# The Case for Structured Random Codes in Network Capacity Theorems

Bobak Nazer and Michael Gastpar

### Abstract

Random coding arguments are the backbone of most channel capacity achievability proofs. In this paper, we show that in their standard form, such arguments are insufficient for proving some network capacity theorems: structured coding arguments, such as random linear or lattice codes, attain higher rates. Historically, structured codes have been studied as a stepping stone to practical constructions. However, Körner and Marton demonstrated their usefulness for capacity theorems through the derivation of the optimal rate region of a distributed functional source coding problem. Here, we use multicasting over finite field and Gaussian multiple-access networks as canonical examples to demonstrate that even if we want to send bits over a network, structured codes succeed where simple random codes fail. Beyond network coding, we also consider distributed computation over noisy channels and a special relay-type problem.

## I. Introduction

Random coding arguments are at the foundation of most channel capacity achievability proofs. The basic idea (for a multiterminal problem) is as follows. Choose several random variables with an appropriate joint distribution. Then, generate high-dimensional codebooks with entries drawn i.i.d. according to this joint distribution. Finally, analyze the error performance of the codebooks in expectation and use this to show the existence of at least one good fixed set of codebooks. This method takes us quite far in network information theory. It has been successfully used to establish the capacity region of the multiple-access channel [1], [2], stochastically degraded broadcast channel [3], and physically degraded relay channel [4], just to name a few. However, an elegant multiterminal problem developed by Körner and Marton showed that purely random code constructions are not always sufficient [5]; structured random codes, such as linear or lattice codes, may be required on the achievability side of the proof. This key insight is the inspiration for this paper.

Structured random codes are usually considered to shed light on issues related to practical constructions. Given a capacity theorem, it is often of interest to demonstrate the existence of a capacity-achieving linear code to show that the codebook size (or complexity) need not be exponential in the blocklength. Linearity also often enables many complexity-saving reductions in decoding algorithms. However, the conditions for structured random codes to be capacity-achieving are often more restrictive than those for unrestricted random codes. For instance, linear codes achieve capacity for point-to-point channels only when the noise is symmetric [6], [7]. Thus, it is tempting to believe that random codes are a strictly more powerful tool for proving capacity theorems. We will show that, in a network setting, structured random codes can be more powerful than purely random codes.

### A. Prior Work

To the best of our knowledge, the first example of a scenario where structured codes outperform random codes was given by Körner and Marton in [5]. In their consideration, a central decoder wants to reconstruct the parity of two correlated sources seen by separate encoders. By using linear codes at the encoders, they were able to access the full rate region. Note that gains are only seen in this problem over random codes when the underlying sources are correlated. In [8], we showed that there are large gains, proportional to the number of users, for computing functions over multiple-access channels with structured codes, regardless of the source dependencies. We applied these results to joint source-channel sensor network scenarios in [9], [10].

Recently, Krithivasan and Pradhan showed that lattice coding results in an improved inner bound for the distributed compression of a linear function of jointly Gaussian sources [11]. In particular, for two positively correlated Gaussian sources, recovering the difference requires a lower sum rate if one uses a lattice code.

For the above scenarios, the computation of a function is directly called for in the problem statement, so structured codes seem to be a natural fit. Somewhat surprisingly, if we are only required to send bits across a network, structured codes can still provide gains. Some instances of this phenomenon, which we term *structural gain*, have been independently studied by several groups. One interesting direction is using lattices for interference cancellation. Philosof, Khisti, Erez, and Zamir demonstrated that for a "dirty" multiple-access channel with two additive interferences, one known non-causally at each encoder, dirty paper coding can be successfully implemented with a lattice code but not with a purely random code [12]. Recently, Cadambe and Jafar [13] proposed the concept of interference alignment. In essence, encoders use their knowledge of the channel to align their transmissions so that for each receiver all interferences lie in one subspace and the desired signal lies on another. For a many-to-one interference channel, Bresler, Parekh and Tse use a lattice to ensure that all the interferences seen at the relevant decoder are aligned [14]. It was also suggested in [15] that structured codes will be useful for implementing interference alignment and similar schemes in general multi-user networks. Finally, Sanderovich, Peleg, and Shamai find a scaling law for distributed interference cancellation using lattices in [16].

Another instance of structural gain is seen in network coding for wireless networks. Instead of avoiding interference from other users, we can use a structured code to compute functions reliably on multiple-access channels. We developed such network coding strategies for discrete channels in [17] and extended our ideas to the Gaussian case in [18]. Narayanan, Wilson, and Sprintson use similar techniques for the two-way relay channel to allow the relay to decode and retransmit only the sum of the transmitted messages [19]. This strategy was extended to the two-way relay channel with unequal transmit powers by Nam, Chung, and Lee [20].

Structured codes are also useful for parallel relay networks with non-white noise spectra. It has been shown that amplify-and-forward is asymptotically optimal for a parallel relay network with white noise and that it significantly outperforms standard random coding strategies [21]. Using a lattice scheme, Kochman, Khina, Erez, and Zamir have managed to attain similar gains for arbitrary noise spectra [22].

### B. Paper Organization

In this paper, we survey several of our recent capacity results based on structured codes to show that even if we simply want to send bits across a network, random coding arguments are not enough.[1] Our main problem formulation is inspired by the recent literature on network coding. Essentially, a single sender must multicast messages to several receivers over a graph of point-to-point and multiple-access channels. We use our recent work on computing functions over multiple-access channels as part of an overall network code. These computation codes are inherently structured and efficiently convert noisy multiple-access channels into reliable computational units [8]. We will also consider a relay-type problem with multiple senders and a single receiver.

The paper is organized as follows. In Section II, we give definitions, particularly for structured codes. We review the best known results for point-to-point channel coding with structured codes in Section III. We describe results on distributed computation using structured codes in Section IV, including the Körner-Marton problem and computation over multiple-access channels. In Section V, we examine a variant of the relay problem. In Section VI, we develop our network coding problem for additive noise multiple-access networks over finite fields as well as the Gaussian case. Finally, in Section VII we discuss open problems and future directions.

## II. DEFINITIONS

We now describe the notions of a purely random code and a structured random code. In later sections, we will show examples of problems for which structured random codes perform well in expectation while purely random codes do not. This enables us to show the existence of at least one good codebook since at least some non-vanishing fraction of the randomly generated codebooks must be good in order to keep the expectation good.

*Definition 1:* Choose an alphabet $\mathcal{X}$. An $(n, R)$ *code*, $\mathcal{C}$, is a set of $2^{nR}$ distinct length-$n$ vectors in $\mathcal{X}^n$. Each vector, $\mathbf{c} \in \mathcal{C}$, is referred to as a *codeword*.

---

[1] A preliminary version of this paper appeared in the 2007 Information Theory Workshop, Lake Tahoe, CA [23].

We now give an informal definition of what we mean by a purely random coding argument. Unfortunately, it is quite difficult to give a formal definition that includes all current techniques (such as superposition coding) but excludes roundabout ways of constructing a set of linear codes by complicated thinning arguments.

A *purely random code* is a code whose codewords are generated according to a distribution such that the elements of each codeword are independent and identically distributed (i.i.d.). For example, for an $(n, R)$ purely random code for a point-to-point channel, we choose a probability distribution function (pdf) $p(x)$ and draw each of the $2^{nR}$ length-$n$ codewords independently according to $p(x^n) = \prod_{i=1}^{n} p(x)$.

The performance of a random code is often evaluated in expectation under a decoding rule such as maximum-likelihood or joint typicality. Given that it performs well in expectation over the random codebook it is then argued that at least one good fixed codebook must exist.

In a multi-user communication problem, more than one codebook is required. Here, it is often useful to allow for some dependence between codebooks but for a purely random code one would still generate them elementwise i.i.d. according to a joint pdf. For instance, for two users whose codebooks each have $2^{nR}$ elements, we can choose a distribution $p(x_1, x_2)$. Then, we draw pairs of codewords independently according to $p(x_1^n, x_2^n) = \prod_{i=1}^{n} p(x_1, x_2)$.

The basic multi-user random coding construction above has been extended to several powerful generalizations in the information theory literature. These include random binning [24], block Markov coding [4], [25], superposition coding [3], [26], and compress-and-forward [4]. With these tools in hand, most of the currently known achievability results of network information theory can be derived. We collectively refer to these strategies as purely random coding as they focus on creating codes whose elements have the desired dependencies but pay no attention to their algebraic structure.

Given the large number of such strategies, it is extremely hard to bound the performance of all possible purely random coding arguments. For the scope of this paper, we limit ourselves to comparisons to the expected performance of the best known purely random coding argument for a particular problem class. For example, for a distributed lossless compression problem (such as the Slepian-Wolf problem [27]), we compare to the performance of random binning in expectation.

We now turn to codes that allow for the design of both the codebook pdf and algebraic structure. A *structured random code* (or structured code) is a code that is randomly generated in a way that ensures a particular algebraic structure. Given two codewords in a structured code, there is a fixed function that can be applied symbolwise such that the resulting vector is also a codeword. For instance, to ensure linearity, we can choose symbolwise addition as the fixed function. As with a purely random code, the performance of a structured code is evaluated in expectation over the codebooks. In the next two subsections, we provide details on two well-studied classes of structured codes: linear codes and lattice codes.

### A. Linear Codes

Linear codes are the most commonly used structured codes as they often enable many complexity reductions in the encoding and decoding algorithms.

*Definition 2:* Let $\mathbb{F}$ be a finite field. An $(n, R)$ linear code, $\mathcal{C}$, is a structured code on $\mathbb{F}^n$ that is closed under the additive operation of $\mathbb{F}$. That is, given any two codewords, $\mathbf{c}_1, \mathbf{c}_2 \in \mathcal{C}$, adding the codewords symbolwise results in another codeword:

$$\mathbf{c}_1 = (x_{11}, x_{12}, \ldots, x_{1n}) \in \mathcal{C} \tag{1}$$

$$\mathbf{c}_2 = (x_{21}, x_{22}, \ldots, x_{2n}) \in \mathcal{C} \tag{2}$$

$$\mathbf{c}^* \triangleq (x_{11} + x_{21}, x_{12} + x_{22}, \ldots, x_{1n} + x_{2n}) \tag{3}$$

$$\implies \mathbf{c}^* \in \mathcal{C} \tag{4}$$

A *random linear code* is just a linear code drawn according to some distribution. For every linear code there is at least one matrix $\mathbf{H} \in \mathbb{F}^{n\hat{R} \times n}$ such that each codeword, $\mathbf{c} \in \mathcal{C} \subset \mathbb{F}^n$, can be written as $\mathbf{x}\mathbf{H}$ for some $\mathbf{x} \in \mathbb{F}^{nR}$.

### B. Lattice Codes

Lattice codes are codes with a linear structure over the reals which are often used as a complexity-reducing code for an AWGN channel. A lattice code is an appropriately chosen subset of a lattice, which is a discrete subgroup of $\mathbb{R}^n$. We cannot simply use a lattice as a codebook on its own as it has infinite extent and, as a result, an infinite number of codewords, even for a finite dimension. Thus, for finite capacity channels, we limit ourselves to a subset of the lattice which is also often chosen to satisfy some type of transmit cost constraint (i.e. the power constraint on the AWGN channel).

*Definition 3:* An $n$-dimensional *lattice*, $\Lambda$, is a set of points in $\mathbb{R}^n$ such that if $\mathbf{x}, \mathbf{y} \in \Lambda$, then $\mathbf{x} + \mathbf{y} \in \Lambda$, and if $\mathbf{x} \in \Lambda$, then $-\mathbf{x} \in \Lambda$. A lattice can always be written in terms of a generator matrix $\mathbf{G} \in \mathbb{R}^{n \times n}$:

$$\Lambda = \{\mathbf{x} = \mathbf{z}\mathbf{G} : \mathbf{z} \in \mathbb{Z}^n\} \tag{5}$$

where $\mathbb{Z}$ represents the integers.

*Definition 4:* A *lattice quantizer* is a map, $Q : \mathbb{R}^n \to \Lambda$, that sends a point, $\mathbf{x}$, to the nearest lattice point in Euclidean distance:

$$\mathbf{x_q} = Q(\mathbf{x}) = \arg \min_{\mathbf{l} \in \Lambda} ||\mathbf{x} - \mathbf{l}||_2 \tag{6}$$

*Definition 5:* Let $[\mathbf{x}] \bmod \Lambda = \mathbf{x} - Q(\mathbf{x})$. The mod $\Lambda$ operation satisfies:

$$[[\mathbf{x}] \bmod \Lambda + \mathbf{y}] \bmod \Lambda = [\mathbf{x} + \mathbf{y}] \bmod \Lambda \quad \forall \mathbf{x}, \mathbf{y} \in \mathbb{R}^n \tag{7}$$

*Definition 6:* The *fundamental Voronoi region*, $\mathcal{V}_0$, of a lattice, is the set of all points that are closest to the zero vector: $\mathcal{V}_0 = \{\mathbf{x} : Q(\mathbf{x}) = \mathbf{0}\}$.

*Definition 7:* An $(n, R)$ lattice code, $\mathcal{C}$, is a code with elements taken from the intersection of some $n$-dimensional lattice $\Lambda$ and a convex $n$-dimensional shape $T$ (which is usually chosen to meet some type of power constraint.)

$$\mathcal{C} = \Lambda \cap T \tag{8}$$

$$|\mathcal{C}| = 2^{nR} \tag{9}$$

*Remark 1:* The constraint on a lattice code is often chosen to be the Voronoi region of another lattice that is nested within the original. For instance, let $\Lambda_1 \subset \Lambda_2$ and let $\mathcal{V}_{0,1}$ be the Voronoi region of $\Lambda_1$. Then, a lattice code $\mathcal{C}$ can be defined as

$$\mathcal{C} = \Lambda_2 \cap \mathcal{V}_{0,1} = \{\mathbf{x} : \mathbf{x} = \mathbf{y} \bmod \Lambda_1, \ \mathbf{y} \in \Lambda_2\}. \tag{10}$$

For more on nested lattice codes and their applications, see [28]. Nested lattices are often employed to enable lattice decoding, where the receiver decodes to the nearest lattice point whether or not it is part of the allowable set of the lattice code.

More broadly, we can consider nonlinear structured codes as well. These codes may prove quite useful in proving their own network capacity theorems, which may even be unavailable to linearly structured codes. However, such codes are beyond the scope of this paper. On a related note, in the network coding literature, linear codes have been shown to attain the same performance as random coding in the multicast case [29], [30]. However, Dougherty, Freiling, and Zeger constructed a (nonmulticast) network for which the linear network coding capacity is significantly lower than the non-linear network coding capacity [31].

A great deal of work has gone into showing that linear and lattice codes are sufficient for many channel coding and source coding problems. In the following section, we will briefly review some of these results.

## III. STRUCTURED CODES FOR POINT-TO-POINT PROBLEMS

We will now review some of the previous work on the existence of capacity-achieving structured codes for classical point-to-point problems.

For additive noise channels, structured codes can achieve rates all the way up to capacity. Work in this area began with Elias' proof that there exist binary linear codes which are good for channel coding over the binary symmetric channel (BSC) [32]. Essentially, the proof shows that a codebook generated from a binary matrix with i.i.d. Bernoulli$(\frac{1}{2})$ entries has pairwise independent codewords that look as if they were drawn elementwise i.i.d. from a Bernoulli$(\frac{1}{2})$ distribution. As the standard random coding achievability proof only requires pairwise independence for the union bound, we can complete the capacity theorem. (See [33, §6.2] for a full proof.)

In fact, Elias' method extends to a much larger class of channels. If we assume the alphabet is over a finite field and the uniform input distribution is capacity-achieving (which is the case for additive noise), then a matrix with elements chosen i.i.d. and uniformly from the finite field will suffice. This is captured in the following lemma from Problem 2.1.11 in [34].

*Lemma 1:* Let $\mathbf{w} \in \mathbb{F}^k$ be the message and let the channel output be given by $\mathbf{y} = \mathbf{x} + \mathbf{z}$ where $\mathbf{x}, \mathbf{y}, \mathbf{z} \in \mathbb{F}^n$ and $\mathbf{z}$ is an i.i.d. sequence. Then the capacity of the channel is given by $C = \log |\mathbb{F}| - H(Z)$ and can be achieved with a linear code $\mathbf{G}^{k \times n}$ so that $\mathbf{x} = \mathbf{w}\mathbf{G}$. Specifically, for any $\epsilon > 0$ and $n$ large enough, $\Pr(c(\mathbf{y}) \neq \mathbf{w}) < \epsilon$ where $c(\cdot)$ is the maximum-likelihood (ML) estimate of $\mathbf{w}$. Note that $k < nC$ is required to stay below the capacity.

Linear codes can also be used for compressing any discrete alphabet source so long as the rate is higher than the source entropy. Moreover, they can reach any point in the Slepian-Wolf rate region for distributed compression [35].

For additive white Gaussian noise (AWGN) channels, showing that lattice codes are sufficient to reach capacity was considerably more challenging. An AWGN point-to-point channel has an output $Y \in \mathbb{R}$ which can be written as $Y = X + Z$ where $X$ is the channel input and $Z$ is i.i.d. Gaussian noise with variance $N$. Unlike the discrete alphabet case, AWGN channel encoders are usually subject to a power constraint of the form $\frac{1}{n}\sum_{i=1}^{n}(x[i])^2 \leq P$. The capacity of an AWGN channel is well-known to be:

$$C = \frac{1}{2}\log\left(1 + \frac{P}{N}\right) \tag{11}$$

Much effort was focused on finding lattices that when intersected with an $n$-dimensional ball of radius $\sqrt{nP}$ centered at 0 form a capacity-achieving code. Urbanke and Rimoldi showed that such lattices exist for minimum angle decoding [36]. Further work by Erez and Zamir showed that nested lattice codes can be capacity-achieving under lattice decoding, i.e. decoding to the closest lattice point whether or not it is an allowable codeword [37].

As in the discrete case, Gaussian source coding can be performed optimally with lattices [38]. Recall that the rate distortion function for mean-squared error for an i.i.d. Gaussian source with variance $\sigma_S^2$ is:

$$R(D) = \frac{1}{2}\log\left(\frac{\sigma_S^2}{D}\right) \tag{12}$$

Erez, Litsyn, and Zamir proved that there exist lattices which are simultaneously good for AWGN channel coding and Gaussian source coding [39]. These will be very useful to us in proving our theorems. Below we give a formal statement on the existence of lattice codes for channel coding and source coding:

*Lemma 2:* Let $\mathbf{s}$ be a length-$n$ Gaussian vector with variance $\sigma_S^2$. There is a sequence of lattices, $\Lambda_n$, such that quantizing $\mathbf{s}$ to the lattice points inside $\Lambda_n \cap B_0(\sqrt{n\sigma_S^2})$ is sufficient for recovery at mean-squared error (or distortion) $D$. Furthermore, the number of points in $\Lambda_n \cap B_0(\sqrt{n\sigma_S^2})$ satisfies:

$$\lim_{n \to \infty} \frac{1}{n}\log\left|\Lambda_n \cap B_0(\sqrt{n\sigma_S^2})\right| = \frac{1}{2}\log\left(\frac{\sigma_S^2}{D}\right) \tag{13}$$

Transmitting points from the same sequence of lattices (up to inflation) intersected with a power sphere $B_0(\sqrt{nP})$ followed by maximum-likelihood decoding is sufficient for achieving any rate below the capacity of an AWGN channel:

$$\lim_{n \to \infty} \frac{1}{n}\log\left|\Lambda_n \cap B_0(\sqrt{nP})\right| = \frac{1}{2}\log\left(1 + \frac{P}{N}\right) \tag{14}$$

where $P$ is the transmit power constraint, $N$ is the noise variance, and the average probability of error goes to 0 as $n \to \infty$.

See [39] for a full proof of the existence of such lattices. Note that their goodness for AWGN capacity under ML decoding essentially follows from Theorem 1 in [40]. There Loeliger shows that lattices from "Construction A" (the employed method of randomly generating lattices) satisfy the Minkowski-Hlawka theorem in expectation. This is all that is needed to invoke the results of Urbanke and Rimoldi for AWGN capacity under ML decoding [36]. In fact, one can establish that the same lattice can be simultaneously good for source coding and channel coding with lattice decoding [37]. Many classical multiterminal problems with a random coding solution also have a lattice coding solution. We refer the interested reader to [28] for an excellent survey of these ideas.

## IV. Distributed Computation

When we are interested in computing a function of distributed sources, structured codes can be extremely useful. In essence, an appropriately chosen structured code will commute with respect to the desired function. The function can then be applied to codewords instead of the original sources. This technique can reduce the required rates as perfect reconstruction of the original sources is no longer required.

We first review the Körner-Marton problem for recovering the parity of two distributed, correlated binary sources [5]. Random binning fails in expectation whereas random *linear* binning succeeds. Next, we will review results that show that large gains are possible for computing functions of sources over noisy multiple-access channels [8].

### A. Körner-Marton Problem

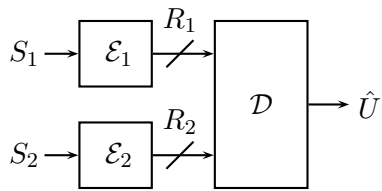

Fig. 1.  Körner-Marton Problem

Let the vector source $(S_1, S_2)$ be generated i.i.d. from the following joint probability distribution function (pdf):

$$\Pr(S_1 = 0, S_2 = 0) \ = \ \Pr(S_1 = 1, S_2 = 1) \ = \ \frac{1-p}{2}$$
$$\Pr(S_1 = 0, S_2 = 1) \ = \ \Pr(S_1 = 1, S_2 = 0) \ = \ \frac{p}{2} \tag{15}$$

A simple calculation will show that $S_1$ and $S_2$ have uniform marginal distributions. We would like to reconstruct the mod-2 sum, $U = S_1 \oplus S_2$, at the decoder with vanishing probability of error. Note that $H(U) = h_B(p)$ where $h_B(p)$ is the usual binary entropy function:

$$h_B(p) = -p \log p - (1-p) \log (1-p) \tag{16}$$

More formally, we would like to find the set of rates $R_1$ and $R_2$ such that there exist two encoders and a decoder:

$$\mathcal{E}_j : \{0,1\}^n \to \{0,1\}^{nR_j} \quad j = 1, 2 \tag{17}$$
$$\mathcal{D} : \{0,1\}^{nR_1} \times \{0,1\}^{nR_2} \to \{0,1\}^n \tag{18}$$

such that the probability of error for recovering $U$ goes to 0 in the blocklength:

$$\hat{\mathbf{u}} = \mathcal{D}\left(\mathcal{E}_1(\mathbf{s}_1), \mathcal{E}_2(\mathbf{s}_2)\right) \tag{19}$$
$$\lim_{n \to \infty} P\left(\hat{\mathbf{u}} \neq \mathbf{u}\right) = 0 \tag{20}$$

#### 1) Optimal Rate Region:

*Theorem 1 (Körner-Marton):* The rate region for distributed compression of $U = S_1 \oplus S_2$ is given by the following constraints:

$$R_1 > h_B(p) \tag{21}$$
$$R_2 > h_B(p). \tag{22}$$

*Proof:* (*Achievability.*) Choose a linear source code, $\mathbf{G} \in \{0,1\}^{n \times nR}$ with rate $R > h_B(p)$ that is sufficient for losslessly compressing $U$. Have each encoder apply this code to its observed source to get $\mathbf{w}_1 = \mathbf{s}_1 \mathbf{G}$ and $\mathbf{w}_2 = \mathbf{s}_2 \mathbf{G}$. These codewords are sent to the decoder which computes $\mathbf{w}_1 \oplus \mathbf{w}_2 = \mathbf{s}_1 \mathbf{G} \oplus \mathbf{s}_2 \mathbf{G} = \mathbf{u} \mathbf{G}$. Since $\mathbf{G}$ was chosen for recovering $U$, decoding is successful.

(*Converse.*) Consider the relaxation where the decoder has full knowledge of $S_2$ and we would like to jointly encode $S_1$ and $U$ to losslessly reconstruct $U$ at the decoder. Note that any scheme that accomplishes this also gives the decoder a lossless reconstruction of $S_1$. Thus, it can be shown that for joint encoding, $R \geq H(S_1, U|S_2) = H(U|S_2) = H(U) = h_B(p)$ is required for a vanishing probability of error. This implies that for separate encoding of $S_1$ and $U$, $R_1 + R_U \geq h_B(p)$. Similarly, we can get that $R_2 + R_U \geq h_B(p)$. Setting $R_U = 0$ gives the desired result. $\square$

*2) Performance of Best Known Random Code:* For random binning with rates satisfying:

$$R_1 > h_B(p) \tag{23}$$

$$R_2 > h_B(p) \tag{24}$$

$$R_1 + R_2 > 1 + h_B(p) \tag{25}$$

it is easy to show that the decoder can reconstruct the source vectors $\mathbf{s}_1$ and $\mathbf{s}_2$ and $\mathbf{u}$ follows by taking the mod-2 sum. We now argue that with random binning the mod-2 sum cannot be recovered at smaller rates. Suppose that $R_1 + R_2 < 1 + h_B(p)$. We can correctly decode the sum with high probability if all typical pairs assigned to a particular pair bin indices have the same mod-2 sum. This ensures that the decoder will not get confused between several possible sums. There are approximately $2^{n(1+h_B(p))}$ typical pairs but there are at most $2^n$ pairs with the same mod-2 sum (even including atypical sequences). Thus, two typical pairs assigned to the same bin indices only have the same mod-2 sum with vanishing probability. As $R_1 + R_2 < 1 + h_B(p)$, we will definitely have many typical pairs assigned to the same bins and these will almost certainly have different mod-2 sums. As a result, we cannot recover the mod-2 sum correctly at the decoder.

It is possible that something more clever than random binning other than linear coding will lead to a successful proof. However, we are not currently aware of any other successful proof techniques.

The Körner-Marton problem demonstrates that there exist problems for which purely random coding arguments are insufficient. However, the gains depend on the source correlations; for independent sources, there is no advantage to linear codes. A similar phenomenon has been discovered for correlated Gaussian sources by Krithivasan and Pradhan [11]. There if the sources are positively correlated and we only demand the difference at the decoder then lattice coding can be helpful. We now examine a problem for which structured random codes give large gains regardless of the underlying source dependencies.

### B. Computation over Multiple-Access Channels

In the standard multiple-access problem, the decoder must recover the messages sent by each encoder. The rate region for this problem was established in [1], [2]. Suppose now that we are only interested in recovering a function of the transmitted messages. If the multiple-access channel (MAC) is simply a deterministic function of its inputs, then clearly it can be used as a reliable computational unit for evaluating that function. However, if noise is also injected, then some form of coding is required to compute reliably.

In [8], we gave a class of strategies for computing functions over noisy MACs. As it turns out, structured codes are an essential part of the code construction. We briefly describe two distributed linear computation problems below and refer the interested reader to [8] for a more comprehensive study. The proofs of both results are reproduced here for completeness.

*1) Discrete Case:* First we consider sending linear functions over a discrete linear MAC.

Let $\mathbb{F}$ be a finite field and let the vector source $(S_1, S_2, \ldots, S_M) \in \mathbb{F}^M$ be generated i.i.d. from some joint pdf. We would like to reconstruct the linear function, $U = \alpha_1 S_1 + \alpha_2 S_2 + \cdots + \alpha_M S_M$, at the decoder with vanishing probability of error. Each source is seen by an encoder with channel input $X_j \in \mathbb{F}$ for $j = 1, 2, \ldots, M$. The channel output of a discrete linear MAC is given by

$$Y = \beta_1 X_1 + \beta_2 X_2 + \ldots + \beta_M X_M + Z \tag{26}$$

where $Z \in \mathbb{F}$ is drawn i.i.d. according to some pdf.

More formally, we would like to find the highest computation rate $R$ such that there exist $M$ encoders and a decoder:

$$\mathcal{E}_j : \mathbb{F}^{nR} \to \mathbb{F}^n \quad j = 1, 2, \ldots, M \tag{27}$$

$$\mathcal{D} : \mathbb{F}^n \to \mathbb{F}^{nR} \tag{28}$$

such that the probability of error for recovering $U$ goes to 0 in the blocklength:

$$\hat{\mathbf{u}} = \mathcal{D}(\mathbf{y}) \tag{29}$$

$$\lim_{n \to \infty} P(\hat{\mathbf{u}} \neq \mathbf{u}) = 0 \tag{30}$$

*Theorem 2:* The capacity for computing $U$ over a discrete linear MAC is given by:

$$C = \frac{\log |\mathbb{F}| - H(Z)}{H(U)} \tag{31}$$

In fact, a similar result holds for transmitting several (possibly correlated) linear functions. See [8] for more details.

*Proof:* (*Achievability.*) Choose a matrix $\mathbf{H}$ that is appropriate for compressing $U$. Similarly, choose a good point-to-point channel coding matrix $\mathbf{G}$ for overcoming noise $Z$. At each encoder we set $\mathbf{x_j} = \beta_j^{-1}\alpha_j \mathbf{s_j HG}$. After the linear operation performed by the channel, we get:

$$\mathbf{y} = \beta_1 \mathbf{x_1} + \beta_2 \mathbf{x_2} + \cdots + \beta_M \mathbf{x_M} + \mathbf{z} \tag{32}$$

$$\mathbf{y} = \mathbf{uHG} + \mathbf{z} \tag{33}$$

Due to $\mathbf{G}$ we can recover from the additive noise and due to $\mathbf{H}$ we can recover $\mathbf{u}$ with a vanishing probability of error.

(*Converse*). For this class of MACs, we can simply allow the encoders to completely collaborate and get a tight upper bound. This reduces our problem to a point-to-point problem and we can invoke the separation theorem to get a converse: $RH(U) \leq \log |\mathbb{F}| - H(Z)$. $\square$

*2) Gaussian Case:* The natural extension of the discrete problem considered above to the continuous case is transmitting the sum of Gaussian sources over a Gaussian MAC at the minimal mean-squared error. When the source and channel bandwidths are equal, uncoded transmission is optimal. However, given more channel uses than source symbols, we would like to continue to use the additive property of the MAC to our advantage. Below we give an achievable scheme and a lower bound for refining the sum over many channel uses.

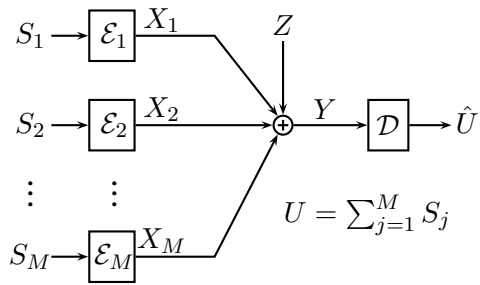

Fig. 2. Reliable Addition over a Gaussian MAC

Each encoder, $\mathcal{E}_j$, sees an i.i.d. Gaussian sequence $\{S_j[i]\}_{i=1}^{k}$ with mean 0 and variance $\sigma_S^2$. For every $k$ source symbols, we are allotted $n = \ell k$ channel uses where $\ell \in \mathbb{Z}_+$.

$$\mathcal{E}_j : \mathbb{R}^k \to \mathbb{R}^n \tag{34}$$

The encoders must satisfy average power constraints:

$$\frac{1}{n}\sum_{i=1}^{n}(x_j[i])^2 \leq P \quad \forall j \in \{1, 2, \ldots, M\} \tag{35}$$

The channel output is just the sum of the channel inputs plus independent Gaussian noise:

$$Y[i] = \sum_{j=1}^{M} X_j[i] + Z[i] \tag{36}$$

where $\{Z[i]\}_{i=1}^{n}$ is an i.i.d. Gaussian sequence with mean 0 and variance $N$.

Our goal is to reconstruct the sum of the sources, $U = S_1 + S_2 + \cdots + S_M$, at the decoder with the lowest possible distortion. Distortion is measured by the usual mean-squared error criterion:

$$D_\ell = \frac{1}{k} \sum_{i=1}^{k} E[(U[i] - \hat{U}[i])^2] \tag{37}$$

*Theorem 3:* Choose $\delta > 0$. For $k$ large enough, the following distortion is achievable for sending $k$ sums of independent Gaussian sources over a Gaussian MAC with $n = \ell k$, $\ell \in \mathbb{Z}_+$ channel uses:

$$D_\ell = M\sigma_S^2 \left( \frac{N}{N + MP} \right) \left( \frac{MN}{N + MP} \right)^{\ell - 1} + \delta. \tag{38}$$

Furthermore, the distortion can be lower bounded by:

$$D_\ell \geq M\sigma_S^2 \left( \frac{N}{N + MP} \right)^\ell. \tag{39}$$

*Proof:* (*Achievability.*) In [41], Kochman and Zamir develop an elegant joint source-channel lattice scheme for sending a Wyner-Ziv Gaussian source over a dirty paper channel. Our distributed refinement scheme consists of two main steps. First, we use uncoded transmission to send a noisy sum to the decoder. Then, we have each encoder run a version of the Kochman-Zamir scheme targeted at the desired sum, $U$. Assume $\ell = 2$. We thus have $2k$ channel uses to convey $k$ sums. We will use the first $k$ channel uses for an uncoded transmission phase. The decoder will then form an MMSE estimate $\hat{U}$ of the sum $U = S_1 + \cdots + S_M$ and use this as side information for the next phase. Thus, $U = Q + \hat{U}$ where $Q$ is an i.i.d. Gaussian sequence with mean 0 and variance $M\sigma_S^2 \frac{N}{N + MP}$.

Choose a sequence of good lattices, $\Lambda_k$, using [39] and scale them such that the normalized second moment of the lattice is $MP$. Let $\mathbf{d}_1, \mathbf{d}_2, \ldots, \mathbf{d}_M$ be independent dither vectors drawn uniformly over the fundamental Voronoi region, $\mathbf{d}_j \sim \text{Unif}(\mathcal{V}_{0,k})$, and made available to the encoders and decoder.

Each encoder transmits $\mathbf{x}_j = \frac{1}{\sqrt{M}} \mathbf{v}_j$ and the decoder receives $\mathbf{y}$:

$$\mathbf{v}_j = [\gamma \mathbf{s}_j + \mathbf{d}_j] \bmod \Lambda_k. \tag{40}$$

$$\mathbf{y} = \frac{1}{\sqrt{M}} \sum_{j=1}^{M} \mathbf{x}_j + \mathbf{z}.$$

The decoder then computes:

$$\mathbf{t} = \alpha \mathbf{y} - \left( \sum_{j=1}^{M} \mathbf{d}_j + \gamma \hat{\mathbf{u}} \right)$$

$$\mathbf{r} = \mathbf{t} \bmod \Lambda_k$$

$$= \left[ \frac{\alpha}{\sqrt{M}} \sum_{j=1}^{M} \mathbf{x}_j + \alpha \mathbf{z} - \sum_{j=1}^{M} (\mathbf{d}_j + \gamma \mathbf{s}_j) + \gamma \mathbf{q} \right] \bmod \Lambda_k$$

$$= \left[ \left( \frac{\alpha}{\sqrt{M}} - 1 \right) \sum_{j=1}^{M} \mathbf{x}_j + \alpha \mathbf{z} + \gamma \mathbf{q} \right] \bmod \Lambda_k.$$

If the second moment of the term inside the modulo operation does not exceed $MP$, the second moment of the lattice, then we can guarantee that:

$$\lim_{k \to \infty} \Pr \left( \mathbf{r} = \left( \frac{\alpha}{\sqrt{M}} - 1 \right) \sum_{j=1}^{M} \mathbf{x}_j + \alpha \mathbf{z} + \gamma \mathbf{q} \right) = 1. \tag{41}$$

See [37] for a detailed discussion of the effect of the dither in this step. The second moment can be controlled by requiring that:

$$\left( \frac{\alpha}{\sqrt{M}} - 1 \right)^2 (M^2 P) + \alpha^2 N + \gamma^2 \sigma_Q^2 \leq MP. \tag{42}$$

This equation will be satisfied by our final choice of the constants $\alpha$ and $\gamma$. The decoder's estimate of the sum is given by:

$$\hat{\hat{\mathbf{u}}} = \beta \mathbf{r} + \hat{\mathbf{u}}$$

$$= \beta \left( \left( \frac{\alpha}{\sqrt{M}} - 1 \right) \sum_{j=1}^{M} \mathbf{x}_j + \alpha \mathbf{z} + \gamma \mathbf{q} \right) + \hat{\mathbf{u}}$$

$$= \beta \left( \left( \frac{\alpha}{\sqrt{M}} - 1 \right) \sum_{j=1}^{M} \mathbf{x}_j + \alpha \mathbf{z} \right) - (1 - \beta\gamma)\mathbf{q} + \mathbf{u}.$$

This estimate gives the following mean-squared error:

$$D = \beta^2 \left( \left( \frac{\alpha}{\sqrt{M}} - 1 \right)^2 M^2 P + \alpha^2 N \right) + (1 - \beta\gamma)^2 \sigma_Q^2.$$

We define the following constants:

$$\alpha = \frac{MP\sqrt{M}}{MP + N}$$

$$\gamma_0 = \sqrt{\frac{MP}{\sigma_Q^2} \left( 1 - \frac{MN}{MP + N} \right)},$$

and let $\gamma \to \gamma_0$ from below as $k \to \infty$. This ensures that Equation (42) is always satisfied. We also set:

$$\beta = \frac{\sigma_Q^2 \gamma}{MP}.$$

As $k \to \infty$, we get that the achieved distortion is:

$$D = M\sigma_S^2 \frac{N}{N + MP} \frac{MN}{N + MP}. \tag{43}$$

This proves the theorem for $\ell = 2$. For all higher values of $\ell$, the scheme can be repeated with the final estimate from the last refinement taken as side information for the next stage. Note that since this works in expectation over the dither random vectors then there must exist fixed vectors which can be used to achieve the same distortion.

(Lower Bound.) Using steps from the converse to the multiple-access problem (see [42, pp. 399-407]) as well as the independence of the sources, we can get that $I(X_1, X_2, \ldots, X_M; Y) \leq \frac{1}{2} \log \left( 1 + \frac{MP}{N} \right)$. It is also clear that the rate distortion function for jointly compressing the sum is given by $R_U(D) = \frac{1}{2} \log \left( \frac{M\sigma_S^2}{D} \right)$. By applying the data processing inequality, we get the desired lower bound. $\square$

The significant gap between the lower bound and the achievable scheme seems due to the distributed nature of the problem. The encoders quantize each source prior to each transmission whereas the lower bound allows for cooperation. Therefore, it seems that the lower bound is quite loose for large $\ell$. We presented a slightly improved achievable scheme as part of [23], [43] but there was an error in the proof.

*3) Performance of Best Known Random Code:* For the problems considered above, random coding arguments perform quite poorly. As in the Körner-Marton problem, the best performance one can hope for is complete reconstruction of the sources at the decoder followed by computing the sum. This reduces the rate at which functions are computed proportionally to the number of users if the sources are independent.

For the discrete problem considered in Section IV-B.1, the random coding argument is as follows. Each encoder generates $2^{nR}$ codewords according to the uniform distribution on $\mathbb{F}$ (which is also the capacity-achieving distribution). Typical sequences are mapped to these codewords. Note that nearly all the probability is concentrated in approximately $2^{nH(S_1, S_2, \ldots, S_M)}$ M-tuples of jointly typical sequences. Of these sequences, approximately $2^{nH(S_1, S_2, \ldots, S_M | U)}$ M-tuples have the same desired function. For a joint typicality decoder, the probability that an incorrect set of vectors is found to be jointly typical with the channel output is approximately $2^{-nI(X_1, X_2, \ldots, X_M; Y)}$. Note that $I(X_1, X_2, \ldots, X_M; Y) = \log |\mathbb{F}| - H(Z)$ for the chosen input distribution. Since $H(S_1, S_2, \ldots, S_M) > H(S_1, S_2, \ldots, S_M | U)$, two sets of M-tuples have the same desired function with vanishing probability. Thus, an incorrect function will be jointly typical with the channel output unless we choose $R$ such that we can distinguish all typical M-tuples:

$$R < \frac{\log |\mathbb{F}| - H(Z)}{H(S_1, S_2, \ldots, S_M)} \tag{44}$$

This is equivalent to requiring complete reconstruction of the sources.

For the Gaussian problem, a similar argument shows that random coding arguments can only achieve the following distortion:

$$D = M\sigma_S^2 \left(\frac{N}{N + MP}\right)^{\ell/M} \tag{45}$$

Again, this is equivalent to reconstructing each source at the decoder using a standard multiple-access technique and then adding them up.

Using structured codes for distributed computation allows several users to exploit the natural function of the channel while overcoming the noise. With a random coding proof, one can only overcome the noise if all of the sources are completely reconstructed at the decoder. As the problem statement includes the evaluation of a function, it seems natural that the codebooks should be structured with respect to the function. It is not so clear that such a property is desirable when we only want to send messages across a network. The problem considered in the next section implies that structured codes have a useful role to play in transmitting bits as well.

## V. Relay-Type Problem

So far we have seen that structured random codes offer gains when we are interested in computing a function in a distributed fashion. We now give a multiterminal channel coding scenario for which structured random codes are better suited to the problem than purely random codes, even though we are only interested in reproducing bits at the decoder.

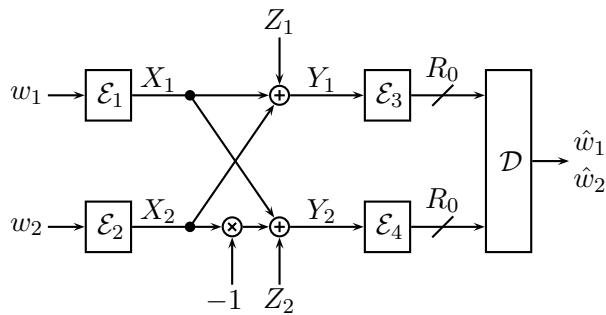

Fig. 3.   Sum-Difference Relay Channel

Two separate users would like to send independent messages $w_1, w_2 \in \{1, 2, \ldots, 2^{nR}\}$ to a decoder across a channel with the help of relays (see Figure 3). Encoder 1 produces channel input $X_1 \in \mathbb{R}$ with average power constraint $P$ and Encoder 2 must produce channel input $X_2 \in \mathbb{R}$ with the same constraint. One relay terminal sees the sum of these signals plus noise: $Y_1 = X_1 + X_2 + Z_1$. The other terminal sees the difference plus noise: $Y_2 = X_1 - X_2 + Z_2$. $Z_1$ and $Z_2$ are independent and Gaussian with variance $N$. Finally each relay has a noiseless bit pipe to the decoder with rate $R_0$. We would like to determine the maximum achievable symmetric rate $R$ to the decoder for a given value of $R_0$.

More formally, we would like to find the maximum $R$ for which there exist four encoders and a decoder:

$$\mathcal{E}_j : \{0, 1\}^{nR} \to \mathbb{R}^n \quad j = 1, 2 \tag{46}$$

$$\mathcal{E}_\ell : \mathbb{R}^n \to \{0, 1\}^{nR_0} \quad \ell = 3, 4 \tag{47}$$

$$\mathcal{D} : \{0, 1\}^{nR_0} \times \{0, 1\}^{nR_0} \to \{0, 1\}^{nR} \times \{0, 1\}^{nR}$$

such that the probability of error for recovering $w_1$ and $w_2$ goes to $0$ in the blocklength:

$$(\hat{w}_1, \hat{w}_2) = \mathcal{D}(\mathbf{y}_3, \mathbf{y}_4) \tag{48}$$

$$\lim_{n \to \infty} P\left((\hat{w}_1, \hat{w}_2) \neq (w_1, w_2)\right) = 0 \tag{49}$$

Note that, ideally, the decoder would have full access to the sum and difference seen at the relays. It could then add and subtract these to get the original signals without interference. Our lattice strategy is geared towards decoding only the sum and difference of the messages at the relays and passing these functions along to the decoder. As seen earlier, implementing such a scheme with random codes is not possible as we effectively need to send the messages in their entirety to get the sum.

### A. Lattice Symmetric Rate

We will use lattices to allow each relay to "compute-and-forward" either the sum or the difference to the decoder. We will proceed by using the scheme for computing the sum of Gaussian sources presented in Section IV-B.2 and then connecting the resulting distortions to transmitting bits. The following lemma will be useful in our analysis.

*Lemma 3:* Assume, for the problem described above, $k$ large enough, and $n = k\ell$, $\ell \in \mathbb{Z}_+$ channel uses, that each user can transmit an i.i.d. Gaussian sequence of length-$k$ with mean $0$ and variance $\sigma_S^2$ ($\mathbf{s}_1$ and $\mathbf{s}_2$) such that the decoder can make estimates $\hat{\mathbf{s}}_1$ and $\hat{\mathbf{s}}_2$, each with mean-squared error (or distortion) $D_{\ell,k}$. Then we can design encoders $\mathcal{E}_j$, $j = 1, 2, 3, 4$ and a decoder $\mathcal{D}$ for transmitting at any rate less than $\frac{1}{2\ell} \log \left( \frac{\sigma_S^2}{D_{\ell,k}} \right)$ from each encoder for any average probability of error greater than zero.

See Appendix A for a full proof.

*Theorem 4:* For the two relay problem described above, the following symmetric rate is achievable using lattice codes:

$$R_{\text{LAT}} = \min \left( \frac{1}{2} \log \left( \frac{1}{2} + \frac{P}{N} \right), R_0 \right). \tag{50}$$

*Proof:* Choose $\mathbf{s}_1$ and $\mathbf{s}_2$ to be i.i.d. length-$k$ Gaussian vectors with mean $0$ and variance $\sigma_S^2$. We will use the channel network $n = k\ell$ times to transmit the sources. Use Theorem 3 to pick a code for sending the sum $\mathbf{u} = \mathbf{s}_1 + \mathbf{s}_2$ to Encoder 3 at distortion $D \triangleq \sigma_S^2 \left( \frac{2N}{N+2P} \right)^{\ell}$. Due to the symmetry of the underlying lattice code and the negative sign on the lower path, Encoder 4 will be able to reconstruct the difference $\mathbf{v} = \mathbf{s}_1 - \mathbf{s}_2$ at distortion $D$ as well.

In order to send the sum and the difference to the final decoder, we will need to requantize them. Pick a Gaussian source code for compressing a variance $2\sigma_S^2$ source to distortion $D_0 = 2\sigma_S^2 2^{-2\ell R_0}$. By the triangle inequality, this requantization step will cause the total distortion for $\mathbf{u}$ and $\mathbf{v}$ to be at most $D_0 + D$. At the decoder we estimate $\mathbf{s}_1$ by $\hat{\mathbf{s}}_1 = \frac{1}{2}(\hat{\mathbf{u}} + \hat{\mathbf{v}})$ and $\mathbf{s}_2$ by $\hat{\mathbf{s}}_2 = \frac{1}{2}(\hat{\mathbf{u}} - \hat{\mathbf{v}})$. It can be checked that these give us the original sources to within distortion $\max \left( 4\sigma_S^2 \left( \frac{2N}{N+2P} \right)^{\ell}, 2D_0 \right)$. Finally, we calculate the rate achievable by each encoder using Lemma 3:

$$R_{\text{LAT}} = \frac{1}{2\ell} \log \left( \frac{\sigma_S^2}{4\sigma_S^2 \max \left( \left( \frac{2N}{N+2P} \right)^{\ell}, 2^{-2\ell R_0} \right)} \right)$$

$$= \min \left( \frac{1}{2} \log \left( \frac{1}{2} + \frac{P}{N} \right), R_0 \right) - \frac{1}{2\ell} \log(4). \tag{51}$$

Thus, for $\ell$ large enough, we can achieve the desired rates. $\square$

The number of refinements, $\ell$, plays an important role in the proof above. Essentially, the sum and difference are "too large" to be sent to the decoder without requantizing them. This requantization doubles the end-to-distortion but since we refine the sum and difference many times, the effect of this doubling is negligible.

We can also achieve the desired rates using a nested lattice code where the relays decode the sum and the difference modulo the coarse lattice. For an example of this type of scheme, see [19] for an application to a two-way relay channel.

A purely random code will not have access to the capacity as we cannot decode the sum and difference without decoding the sources in their entirety. Below we describe the best known random coding strategy for such a problem and demonstrate that it is indeed below the capacity.

### B. Performance of Best Known Random Code

For this type of problem, the best known strategies are decode-and-forward and compress-and-forward. For decode-and-forward, the relays recover the message intended for the decoder and pass it along. For compress-and-forward, the relays quantize their observed signals according to Gaussian codebooks and forward these to the decoder.

*Theorem 5:* The best achievable rates for decode-and-forward and compress-and-forward with Gaussian codebooks are given below:

$$R_{\mathrm{DF}} = \min\left(\frac{1}{4}\log\left(1 + \frac{2P}{N}\right), R_0\right) \tag{52}$$

$$R_{\mathrm{CF}} = \frac{1}{2}\log\left(1 + \frac{2P(2^{2R_0} - 1)}{2P + N2^{2R_0}}\right) \tag{53}$$

*Proof:* For the decode-and-forward scheme, the message from Encoder 1 is meant to be decoded at Encoder 3 and the message from Encoder 2 is meant to be decoded at Encoder 4. However, as Encoders 3 and 4 see the outputs of statistically equivalent multiple-access channels, they can always decode the message intended for the other terminal as well. Thus, we can use the standard multiple-access result to bound the rates and we get that the maximum symmetric rate is $\frac{1}{4}\log\left(1 + \frac{2P}{N}\right)$ (assuming the relay links to the decoder are not the bottleneck). For the compress-and-forward scheme, we can think of quantizing the signals as simply adding additional noise to the observed signals. We can show that this additional noise has variance $\tilde{N} = \frac{2P+N}{2^{2R_0}-1}$. Adding the two received signals together gives an estimate of the first transmitted message with SNR $\frac{4P}{2N+2\tilde{N}}$. Taking the difference gives a similar estimate of the second user's message. $\square$

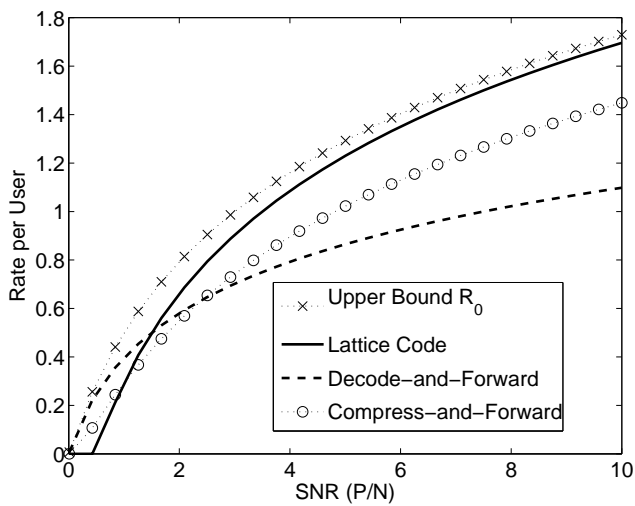

Fig. 4.   Achievable Symmetric Rate Points for Sum-Difference Relay Channel

As an example, we let $R_0 = \frac{1}{2}\log\left(1 + \frac{P}{N}\right)$ and we evaluate the performance of each scheme in Figure 4. The lattice coding scheme is superior to both schemes for SNRs higher than 1.5.

The decode-and-forward scheme completely ignores the benefits of using the sum and the difference and suffers as a result. Although the compress-and-forward scheme lets the decoder see a version of the sum and the difference, this version also includes the noise as we naively forwarded the resulting signal. (Note that if $R_0$ tends to infinity, compress-and-forward is optimal.) By using a lattice, we can just "compute-and-forward" the desired functions.

Most work in multiple-access theory focuses on how many bits can be sent from a set of users to a single receivers. This example indicates that for AWGN networks it may also be useful to characterize exactly how many linear functions can be sent per channel use. This question is partially answered by Theorem 3. In the next section, we will see that converting an AWGN network into a system of linear equations can be fruitful.

## VI. NETWORK CODING WITH MACS

Recently, there has been a great deal of interest in network coding [29], [30], [44]. The key idea is that for multicasting from one sender to multiple receivers over a network of point-to-point channels, the encoders should sometimes only send a function of their incoming messages (i.e. routing is suboptimal). If the network now includes MACs, it is clear that we should try to use these MACs to compute functions of incoming messages when required.

As seen in the previous section, it may be useful to think about multiple-access links in terms of how many functions can be reliably transmitted across them instead of just the standard multiple-access rate region. As detailed in Section IV, structured codes are a natural way of constructing codes for distributed computation.

A channel network is usually thought of as a graph where the vertices are the encoders and decoders and the edges are the point-to-point channels with known capacities. For our problem, we will need a bit more notation to cleanly represent both point-to-point channels and MACs.

*Definition 8:* A *multiple-access network*, $\mathcal{G}_{\text{MAC}}$, consists of the following elements:

1) $\mathcal{V}_N$: the encoder/decoder nodes of the network. Each node, $v$, has a unique label taken from the positive integers, $v \in \mathbb{Z}_+$, and consists of a decoding function $g_{v_j v}$ for each incoming edge $(v_j, v)$ and an encoding function $f_{v v_k}$ for each outgoing edge $(v, v_k)$.
2) $v^S$: the source node. One element of $\mathcal{V}_N$. The source transmits the message, $w \in \{1, 2, \ldots, 2^{nR}\}$.
3) $(v_1^R, v_2^R, \ldots, v_L^R)$: the receiver nodes. Each one is an element of $\mathcal{V}_N$ and produces an estimate of the transmitted message, $\hat{w}_\ell$.
4) $\mathcal{V}_{\text{MAC}}$: the MACs in the network. Each MAC, $m$, has a unique integer label, $m \in \mathbb{Z}_+$. Each MAC has a noise variance $N_m \in \mathbb{R}_+$.
5) $\mathcal{E}_{NN}$: the directed point-to-point channels in the network. Each channel has a unique integer label, $e_{NN} \in \mathbb{Z}_+$, and the labels of its inputs and output nodes are given by the functions $v_{\text{IN}}(e_{NN})$ and $v_{\text{OUT}}(e_{NN})$ respectively. The noise variance for the channel is given by $N_e \in \mathbb{R}_+$.
6) $\mathcal{E}_{NM}$: the input edges from nodes to MACs. Each edge has a unique integer label, $e_{NM} \in \mathbb{Z}_+$, and the labels of its inputting node and destination MAC are given by the functions $v_{\text{IN}}(e_{NM})$ and $v_{\text{OUT}}(e_{NM})$ respectively.
7) $\mathcal{E}_{MN}$: the output edges from a MAC to a node. We assume that the output of a given MAC is only observed by a single node. Each edge has a unique integer label, $e_{MN} \in \mathbb{Z}_+$, and the label of its MAC and destination node are given by $v_{\text{IN}}(e_{MN})$ and $v_{\text{OUT}}(e_{MN})$ respectively.
8) $X_{v_j v_k}[i]$: the channel input on the edge $(v_j, v_k)$ at time $i$. The encoders are constrained to only produce channel inputs from time $i = 1$ to time $i = n$.
9) $Y_{v_j v_k}[i]$: the channel output on the edge $(v_j, v_k)$ at time $i$.

We also assume that there are a finite number of nodes and channels in the network, $|\mathcal{V}_N| + |\mathcal{V}_{\text{MAC}}| + |\mathcal{E}_{NN}| + |\mathcal{E}_{NM}| + |\mathcal{E}_{MN}| < \infty$.

*Definition 9:* A multicast rate, $R$, is *achievable* if $\forall \epsilon > 0$ and $n$ large enough there exist encoding and decoding functions for the network such that the average probability of error is less than $\epsilon$:

$$\hat{w}_\ell = f_{v_\ell^R}(Y_{v_\ell^R}^n)$$
$$\Pr\left(\{\hat{w}_1 \neq w\} \cup \cdots \cup \{\hat{w}_L \neq w\}\right) < \epsilon, \tag{54}$$

where $w \in \{1, 2, \ldots, 2^{nR}\}$ and $Y_{v_\ell^R}^n$ represents all the channel outputs observed by the $\ell^{\text{th}}$ receiver.

*Definition 10:* The *multicast capacity* is the supremum of all achievable multicast rates.

*Definition 11:* A *point-to-point network*, $\mathcal{G}_{\text{POINT}} = (\mathcal{V}_N, \mathcal{E}_{NN})$, is just a multiple-access network without any multiple-access nodes, $\mathcal{V}_{\text{MAC}} = \mathcal{E}_{NM} = \mathcal{E}_{MN} = \emptyset$.

*Definition 12:* A *unit bit pipe network*, $\mathcal{G}_{\text{PIPE}} = (\mathcal{V}, \mathcal{E})$, is just a point-to-point network except all of the channels, $\mathcal{E}$, are taken to be noiseless bit pipes with unit capacity. The encoding/decoding nodes are given by the set $\mathcal{V}$.

Our scheme will give achievable rates for multiple-access networks comprised of either discrete linear or Gaussian MACs. We express the achievable rate through a new point-to-point network that results from an appropriate transformation of our original network. The achievable rate is then given by the multicast capacity of the point-to-point network. We will also demonstrate that in some cases our achievable rates coincide with the simple upper bound due to the max-flow min-cut theorem of Ford and Fulkerson [45].

We now briefly review some results for multicasting over point-to-point channel networks. In [44], it was shown that for a unit bit pipe network the multicast capacity is given by the max-flow min-cut theorem. For each receiver, calculate the maximum information flow across all cuts that separate the source node from that receiver. The multicast capacity is the minimum of all these max-flow values taken over all cuts and receivers. In [29] and [30], it was shown that linear encoding and decoding over a finite field is sufficient to achieve the multicast capacity. Bounds are also given on the required field size. It was independently and concurrently shown by Ho et al. in [46], Jaggi et al. in [47], and Sanders et al. [48] that the field size only needs to be larger than the number of receivers. We reproduce the version from [46] below as it will be useful to us in proving our main theorems.

*Definition 13:* Let $\mathcal{G}_{\mathrm{PIPE}} = (\mathcal{V}, \mathcal{E})$ and let $\mathbb{F}_q$ be a finite field of size $q$. An *algebraic network code* is a set of linear functions for a unit bit pipe network. Specifically, the encoding function from node $v_j$ to node $v_k$ is constrained to be a linear function of its observations from each incoming edge:

$$X_{v_j v_k}[i] = \sum_{v_r} \alpha_{v_r v_j} Y_{v_r v_j}[i]. \tag{55}$$

where $Y_{v_r v_j}[i]$ is the value seen by node $v_j$ at time $i$ on the incoming edge from node $v_r$ and $Y_{v_r v_j}[i], \alpha_{v_r v_j} \in \mathbb{F}_q$ for all $v_r \in \mathcal{V}$.

*Lemma 4 (Ho et al.):* Let $\mathcal{G} = (\mathcal{V}, \mathcal{E})$ be a unit bit pipe network with a single source and $L$ receivers. The multicast capacity is given by the max-flow min-cut bound and can be achieved by an algebraic network code over any finite field larger than $L$ ($\mathbb{F}_q$, $q > L$).

For a full proof, see [49].

Now we need to map these linear functions onto the multiple-access channels in the network. We first explore two examples that are a variant of the butterfly network given in [44]. Then, we state our theorems on multicasting over general finite field and AWGN multiple-access networks.

### A. Motivating Examples

We now look at two variants of the butterfly network from [44, Figure 7] to demonstrate our coding idea. In the original butterfly, recall that it is optimal to send one bit down the left path, a different bit down the right path, and the mod-2 sum of these bits down the center path. In our examples, the center path now includes a multiple-access channel. Our goal will be to add together two bitstreams on the MAC. The first example examines at the binary alphabet case and the second looks at the Gaussian case.

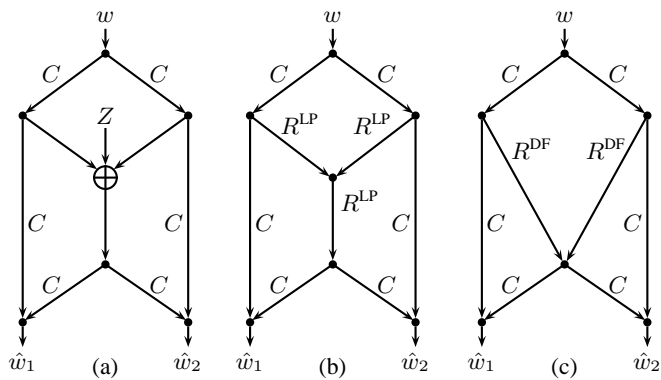

Fig. 5. (a) Binary multiple-access variant of the butterfly network. The MAC in the center is a noisy mod-2 adder with i.i.d. Bernoulli($p$) noise $Z$ (b) Using a linear code, we can achieve the multicast capacity of the MAC butterfly network which is equivalent to the multicast capacity of this transformed network. Here, $R^{\mathrm{LP}} = 1 - h_B(p)$. (c) With a decode-and-forward strategy, we can only achieve rates on the original network that are achievable on this network. Here, $R^{\mathrm{DF}} = \frac{1}{2}(1 - h_B(p))$

*1) Mod-2 Adder MAC:* Consider the channel network in Figure 6(a). Each vertex on the graph represents a decoder/encoder pair. The sender is at the top of the graph and the two receivers are at the bottom. The labeled edges represent noiseless bit pipes each with capacity C. At the center of the graph is a MAC with inputs $X_1$ (from the left) and $X_2$ (from the right) and output $Y = X_1 \oplus X_2 \oplus Z$ where $Z$ is an i.i.d. Bernoulli($p$) sequence. Note that the sum rate of this MAC is upper bounded by $1 - h_B(p)$.

*Theorem 6:* For the channel graph from Figure 5 (a) the multicast capacity is:

$$R = C + \min\left(C, 1 - h_B(p)\right) \tag{56}$$

*Proof:* (*Converse.*) Applying the cutset bound gives that the rate to each receiver is upper bounded by:

$$R \le C + \min\left(C, 1 - h_B(p)\right) \tag{57}$$

(*Achievability.*) We have a block **b** of $n(C + \min\left(C, 1 - h_B(p)\right))$ bits to transmit to both receivers. We will break up **b** in two ways. For the first, we write $\mathbf{b} = [\mathbf{b}_{11}\mathbf{b}_{12}]$ where the first chunk is of length $nC$ and the second is of length $n(\min\left(C, 1 - h_B(p)\right))$. For the second, we write $\mathbf{b} = [\mathbf{b}_{21}\mathbf{b}_{22}]$ where the first chunk is of length $n(\min\left(C, 1 - h_B(p)\right))$ and the second is of length $nC$. We transmit $\mathbf{b}_{11}$ down the left path and $\mathbf{b}_{22}$ down the right path. From $\mathbf{b}_{11}$ we automatically know $\mathbf{b}_{21}$ and from $\mathbf{b}_{22}$ we know $\mathbf{b}_{12}$. We send the mod-2 sum $\mathbf{u} = \mathbf{b}_{21} \oplus \mathbf{b}_{12}$ reliably across the MAC using the linear code from Theorem 2. Finally, this mod-2 sum is conveyed to the receivers. The left receiver can compute $\mathbf{b}_{12} = \mathbf{u} \oplus \mathbf{b}_{21}$ and the right receiver can compute $\mathbf{b}_{21} = \mathbf{u} \oplus \mathbf{b}_{12}$ to fully recover **b**. $\square$

Standard random coding arguments cannot attain the optimal performance over the network in Figure 5 (a). The decode-and-forward and compress-and-forward rates are given by:

$$R_{\text{DF}} = C + \min\left(C, \frac{1 - h_B(p)}{2}\right) \tag{58}$$

$$R_{\text{CF}} = C + \min\left(C(1 - h_B(p)), (1 - h_B(p))\right) \tag{59}$$

If the capacity $C$ of the point-to-point links is small enough, then fully decoding the input messages to the MAC is sufficient. If the capacity $C$ is large enough, forwarding the output of the MAC is sufficient. However, in the intermediate regime, structured codes are necessary.

*2) Gaussian MAC:* Consider the AWGN channel network in Figure 6 (a). Each vertex on the graph represents a decoder/encoder pair. The sender is at the top of the graph and the two receivers are at the bottom. All encoders must satisfy an average power constraint, $\frac{1}{n}\sum_{i=1}^{n} x_j[i]^2 \le P$. The $Z_m$, $m = 1, 2, \ldots, 7$ are drawn i.i.d. according to a Gaussian distribution with mean $0$ and variance $N$.

Again, it is convenient to consider sending Gaussian vectors across the network and then link the performance back to sending bits reliably.

*Lemma 5:* For a given Gaussian multiple-access network, assume for $k$ large enough and $n = k\ell$, $\ell \in \mathbb{Z}_+$ channel uses, that the sender can transmit $B$ i.i.d. Gaussian sequences of length-$k$ with mean $0$ and variance $\sigma_S^2$ such that each receiver can make an estimate of each source sequence to within distortion $D$. Then we can design encoders and decoders for multicasting at any rate less than $\frac{B}{2\ell}\log\left(\frac{\sigma_S^2}{D}\right)$ for any average probability of error greater than zero.

See Appendix A for a full proof.

*Theorem 7:* The following multicast rate is achievable on the channel network in Figure 6(b) using a structured code:

$$R = \frac{1}{2}\log\left(1 + \frac{P}{N}\right) + \frac{1}{2}\log\left(\frac{1}{2} + \frac{P}{N}\right) \tag{60}$$

*Proof:* Choose $\mathbf{s}_1$ and $\mathbf{s}_2$ to be i.i.d. length-$k$ Gaussian vectors with mean $0$ and variance $\sigma_S^2$. We will use the channel network $n = k\ell$ times to transmit the sources where $\ell \in \mathbb{Z}_+$ will be specified later. Use Theorem 3 to pick a code for sending the sum $\mathbf{u} = \mathbf{s}_1 + \mathbf{s}_2$ across the MAC at distortion $D \triangleq 2\sigma_S^2\left(\frac{2N}{N+2P}\right)^\ell$.

We transmit $\mathbf{s}_1$ down the left path and $\mathbf{s}_2$ down the right path at distortion $D$. This requires rate $\frac{1}{2}\log\left(\frac{1}{2} + \frac{P}{N}\right)$. We also transmit the sum of $\mathbf{s}_1$ and $\mathbf{s}_2$ down the MAC and requantize it to distortion $2D$ for transmission to the receivers.

At the left receiver we have estimates $\hat{\mathbf{u}}$ and $\hat{\mathbf{s}}_1$ and we form an estimate $\hat{\mathbf{s}}_2 = \hat{\mathbf{u}} - \hat{\mathbf{s}}_1$. The left receiver knows the source vectors at distortions $\frac{D}{2}$ and $\frac{9D}{2}$, respectively. (The distortions for the right receiver follow by analogy.)

Using Lemma 5 and taking the worst distortion for each source, our achievable rate is given by:

$$R = \frac{1}{\ell} \log \left( \frac{\sigma_S^2}{9 \sigma_S^2 \left( \frac{2N}{N+2P} \right)^\ell} \right) \tag{61}$$

$$= \log \left( \frac{1}{2} + \frac{P}{N} \right) - \frac{1}{\ell} \log (9). \tag{62}$$

Note that we still have $\frac{1}{2} \log \left( 1 + \frac{P}{N} \right) - \frac{1}{2} \log \left( \frac{1}{2} + \frac{P}{N} \right)$ rate on both the left and right paths. We use this to send a common message. Finally, we get that for $\ell$ large enough, we can achieve the desired rate. $\square$

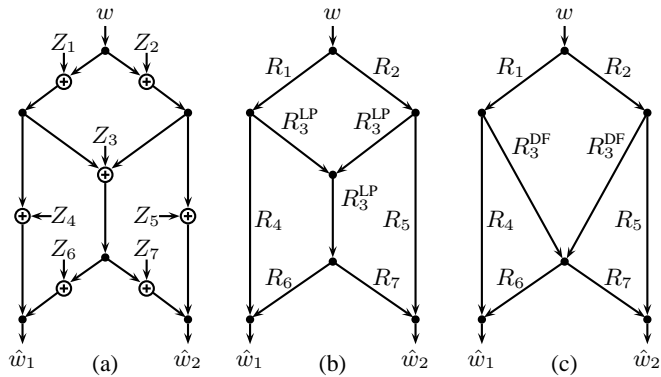

Fig. 6. (a) Gaussian multiple-access variant of the butterfly network. (b) Using a lattice-based code, we can achieve any rate on the MAC butterfly network that is achievable on this transformed network. Here, $R_3^{\mathrm{LP}} = \frac{1}{2} \log \left( \frac{1}{2} + \frac{P}{N_3} \right)$ where $P$ is the per user power of the MAC. (c) With a decode-and-forward strategy, we can only achieve rates on this network where $R_3^{\mathrm{DF}} = \frac{1}{4} \log \left( 1 + \frac{2P}{N_3} \right)$

As in the discrete case, random coding arguments will not suffice for attaining this performance. The best strategies are analogous to that developed in Theorem 5. The encoder following the MAC either decodes the messages in their entirety or quantizes the observed signal and forwards it. Below we give the best achievable rates for decode-and-forward and compress-and-forward with Gaussian codebooks:

$$R_{\mathrm{DF}} = \frac{1}{2} \log \left( 1 + \frac{P}{N} \right) + \frac{1}{4} \log \left( 1 + \frac{2P}{N} \right)$$

$$R_{\mathrm{CF}} = \frac{1}{2} \log \left( 1 + \frac{P}{N} \right) + \frac{1}{2} \log \left( 1 + \frac{P}{N} \left( \frac{P}{3P+N} \right) \right)$$

### B. General Networks

We now give two multicasting results for general multiple-access networks. First, we give the multicast capacity for when the MACs are constrained to be noisy linear functions over a finite field. Second, we give achievable rates for any Gaussian multiple-access network.

*1) Finite Field Multiple-Access Networks:* We assume all of the MACs in our channel network $\mathcal{G}_{\mathrm{MAC}}$ are constrained to be of the form $Y = \alpha_1 X_1 + \alpha_2 X_2 + \dots \alpha_K X_K + Z$ where $\alpha_i \in \mathbb{F}_q \setminus \{0\}$, $X_i \in \mathbb{F}_q$, and $Z$ is an i.i.d. random variable on the alphabet $\mathbb{F}_q$.

*Theorem 8:* The max-flow min-cut bound for multicasting is achievable for a finite field multiple-access network, $G$, if $q > L$, where $q$ is the MAC field size and $L$ is the number of receivers in the network.

By using Theorem 2, we can transform all the MACs in the network into reliable adders. They can then be made part of an overall network code chosen using Lemma 4. For a detailed proof, see [8].

*2) Gaussian Multiple-Access Networks:* We now assume that all channels are either AWGN point-to-point channels or Gaussian MACs. We further assume each user faces an identical power constraint and that the channel quality is controlled by the noise variance $N_m \in \mathbb{R}_+$.

More formally, *all* channels in our channel network $\mathcal{G}_{\mathrm{MAC}}$ are constrained to be of the form $Y_m = X_1 + X_2 + \ldots X_J + Z_m$ where $Z_m$ is an i.i.d. Gaussian sequence of mean $0$ and variance $N_m$. Furthermore, the $X_j$ must satisfy power constraints of the form $\frac{1}{n} \sum_{i=1}^{n} (x_j[i])^2 \leq P$.

The naive approach to a Gaussian multiple-access network would be to first convert all channels, including MACs, into bit pipes using standard random coding arguments. Network coding could then be performed over the resulting bit pipes. This ignores the potential of the MACs to compute as part of an overall network code. Another approach would be to transmit uncoded and use the MACs as noisy linear adders. However, this allows noise to build up unnecessarily as our signals propagate to the receivers. Our strategy uses a lattice code to reliably compute linear functions over the MACs in the network. This effectively converts the MACs into linear processing units and reduces the network to a point-to-point network for which multicasting is well understood (see Figure 7 for an example). The key question is what rates should we assign to the incoming and outgoing links to these nodes in our reduction. We will call this the *linear processing rate* and we will demonstrate that it is at least:

$$R_{LP} = \frac{1}{2} \log \left( \frac{1}{J_m} + \frac{P}{N_m} \right) \tag{63}$$

where $N_m$ is the noise variance associated with that MAC and $J_m$ is the number of inputs to the MAC.

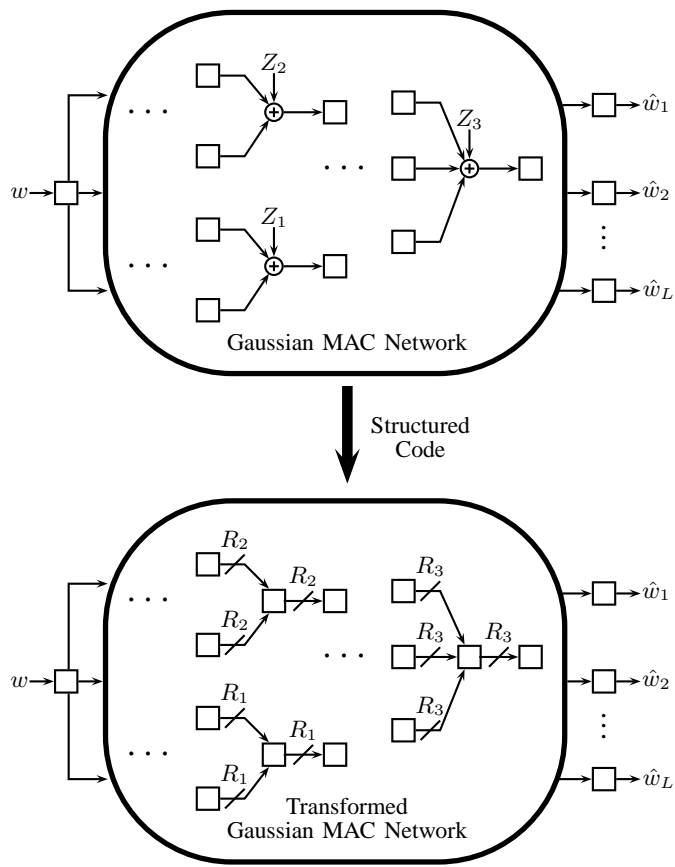

$$R_m = \frac{1}{2} \log \left( \frac{1}{J_m} + \frac{P}{N_m} \right)$$

Fig. 7. With a structured code we can convert every MAC in a network into a reliable linear relay with incoming and outgoing rate links given by the linear processing rate. Note that $J_m$ is the number of inputs to the MAC (2 for the MACs on the left and 3 for the MAC on the right) and $N_m$ is the noise variance for MAC $m$.

We now give a formal statement of our result.

*Definition 14:* Let $\mathcal{G}_{\text{MAC}}$ be a Gaussian multiple-access network. An equivalent point-to-point network, $\mathcal{G}' = (\mathcal{V}', \mathcal{E}')$, is constructed from the original network by the following steps:

- Let the set of encoder/decoder nodes, $\mathcal{V}'$, in the new network be given by the original encoder/decoder nodes as well as the original MACs, $\mathcal{V}' = \mathcal{V}_N \cup \mathcal{V}_{\text{MAC}}$.
- Let the channels in the new network, $\mathcal{E}'$, be given by the original point-to-point channels as well as the input and output edges to the MACs, $\mathcal{E}' = \mathcal{E}_{NN} \cup \mathcal{E}_{NM} \cup \mathcal{E}_{MN}$. The connectivity of these edges is the same as in the original network.
- Set the capacity of the edges taken from $\mathcal{E}_{NM}$ and $\mathcal{E}_{MN}$ to be the linear processing rate, $R_{LP}$.

*Theorem 9:* Let $\mathcal{G}_{\text{MAC}}$ be a Gaussian multiple-access network. Any multicast rate achievable on the equivalent point-to-point network is achievable on the original network for the following linear processing rate:

$$R_{LP} = \frac{1}{2} \log \left( \frac{1}{J_m} + \frac{P}{N_m} \right) \tag{64}$$

where $N_m$ is the noise variance of the MAC $m$ associated to these edges in the original network and $J_m$ is the number of inputs to these MACs.

See Appendix B for a full proof.

Using a purely random code, we can also achieve the same network transformation but with a different linear processing rate given by:

$$R_{DF} = \frac{1}{2J_m} \log \left( 1 + \frac{J_m P}{N_m} \right). \tag{65}$$

For sufficiently high SNR, the performance of the structured random code is superior.

## VII. Discussion and Open Problems

In this paper, we have examined several interesting problem classes where random coding arguments seem insufficient. Surprisingly, for some of these problems, structured random codes are enough to close the gap and determine a tight lower bound to capacity. These structured codes were used to allow certain decoders in the network to recover a function of messages reliably. These functions could then be passed along to a final decoder which, equipped with a full rank set of linear equations, could decode the individual messages. Such a phenomenon has already been well-characterized in the field of network coding. However, in the standard network coding problem, messages are collected at an encoder and then their function is computed. For a graph of point-to-point channels, this is optimal but if the graph also includes multiple-access channels, structured codes seem necessary to convert these into reliable computational units as well.

Our definition of a purely random code is certainly debatable. Still, it seems to us that any code that is capable of recovering these results has some form of structure, whether or not it is generated in the same manner as our codes. We close by listing some open problems around the topic of structured random codes and AWGN network capacity.

*Open Problem 1:* All of the structured codes considered in this paper were linear. It would be interesting to find non-linear structured codes and a canonical problem class where they outperform purely random codes.

*Open Problem 2:* A proof that if one achieves the maximum rate in one of the problems considered in this paper then almost all codewords of the underlying code belong to the same linear subspace. This seems most straightforward for the Körner-Marton problem or computation over MACs.

*Open Problem 3:* A tight characterization of the best achievable distortion for Gaussian computation over MACs (see Theorem 3). We currently suspect that the lower bound is quite loose for large $\ell$.

*Open Problem 4:* Determine the multicast capacity for Gaussian multiple-access networks.

*Open Problem 5:* Find a transform from an arbitrary AWGN network capacity problem (multiple senders and receivers) into a system of linear of equations and find achievable rates using lattices. Is this an optimal strategy in general?

## Appendix I
## Achievable Rates from Distortions

We now show that if we can transmit an i.i.d. Gaussian sequence from a transmitter to one or more receivers at a given mean-squared error then we can also communicate bits.

*Theorem 10:* Let $\mathbf{s}$ be a length-$k$ i.i.d. Gaussian vector with mean zero and variance $\sigma_S^2$ available at a transmitter. Assume that we have a coding scheme that uses $n = k\ell$ channel uses, $k, \ell \in \mathbb{Z}_+$, to send $\mathbf{s}_j$ to a subset $G \subset \{1, 2, \ldots, L\}$ of the receivers with mean squared error at most $D_{\ell,k}$. Then, there exists a coding scheme that can send a message $w$ from the same transmitter to all receivers in $G$ at any rate $R < \frac{1}{2\ell} \log\left(\frac{\sigma_s^2}{D_{\ell,k}}\right)$ for any average probability of error greater than zero.

*Proof*: Fix $k$ and $\ell$. Choose the encoders and decoder in the network such that we achieve the specified distortions at the receivers. We have by the data processing inequality:

$$\frac{k}{2} \log\left(\frac{\sigma_S^2}{D_{\ell,k}}\right) \leq I(S^k; \hat{S}_\ell^k) \tag{66}$$

$$\leq I(X^n; Y_g^n). \tag{67}$$

Thus we know that there exists a multiletter input distribution, $p(x^n)$, such that the mutual information to each receiver is lower bounded by the left-hand side of (66). We now define supersymbols of length $n$:

$$\tilde{X}[i] = [X[(i-1)n + 1], X[(i-1)n + 2], \cdots, X[in]]$$
$$\tilde{Y}_\ell[i] = [Y_\ell[(i-1)n + 1], Y_\ell[(i-1)n + 2], \cdots, Y_\ell[in]].$$

The supersymbols $\tilde{X}$ and $\tilde{Y}$ take values in the alphabets $\tilde{\mathcal{X}} = \mathbb{R}^n$ and $\tilde{\mathcal{Y}} = \mathbb{R}^n$ respectively.

Keep all encoders and decoders in the network the same as in the distortion-achieving case except for those at the source and the receivers, $v_S, v_g^R$, $\forall g \in G$. Thus, $p(\tilde{y}_g | \tilde{x})$ is a memoryless channel. Generate a random codebook with $2^{NR}$ length $N$ codewords (one for each message in $\{1, 2, \ldots, 2^{NR}\}$) with each symbol drawn i.i.d. from $\tilde{\mathcal{X}}$ for some $N \in \mathbb{Z}_+$ and $R > 0$. The $g^{\text{th}}$ receiver upon seeing $\tilde{Y}_g^N$ uses a maximum likelihood rule to infer the original message $w$. Denote this estimate by $\hat{w}_g$. It follows from [50] that for such a channel and $N$ large enough, the average probability of error over codebooks and messages for receiver $g$, $\bar{\text{Pr}}(\hat{W}_g \neq W)$ can be made less than $\frac{\epsilon}{L}$ if $R < I(\tilde{X}; \tilde{Y})$.

It follows from the union bound that the probability that any receiver is in error averaged over all codebooks and messages satisfies:

$$\bar{P}_e \leq \sum_{g \in G} \bar{\text{Pr}}(\hat{w}_g \neq w) < \epsilon. \tag{68}$$

Finally, we get that there exists at least one fixed codebook with average probability of error at most $\bar{P}_e$, otherwise the average over all codebooks would not hold. This completes the proof. $\square$

## Appendix II
## Gaussian Network Coding Proof

In this appendix, we provide a full proof of our Gaussian multiple-access network coding theorem for completeness. First, we show how to compute a linear function of Gaussian sources over a Gaussian MAC at the desired linear processing rate.

*Lemma 6:* Let $S_1, S_2, \ldots, S_J$ be Gaussian sources with variances $\sigma_1^2, \sigma_2^2, \ldots, \sigma_J^2$ respectively. Let $\sigma_{\text{MAX}}^2 = \max_j \sigma_j^2$. Each source is seen at one encoder with power $P$ which faces a Gaussian MAC with noise variance $N$. Let $q$ be a positive prime number and let $U = \beta_1 S_1 + \beta_2 S_2 + \cdots + \beta_J S_J$ where $\beta_j \in \{0, 1, 2, \ldots, q-1\}$. Then given $k$ vectors of source symbols and $n$ channel uses where $n = k\ell$, the decoder can make an estimate of $U$ at distortion

$$D_\ell = J(q-1)^2 \sigma_{\text{MAX}}^2 \left(\frac{JN}{N+P}\right)^\ell. \tag{69}$$

*Proof:* The bulk of the work is done by Theorem 3. To send a linear function, we choose a lattice, $\Lambda$, for use in refining a sum with Theorem 3 as if all the sources had variance $(q-1)^2 \sigma_{\text{MAX}}^2$. At each terminal, we quantize $T_j = \beta_j S_j + W_j$ onto the lattice $\Lambda$ where $W_j$ is an i.i.d. Gaussian random variable available as common randomness to both the encoder and the decoder. Its variance is chosen such that the variance of $T_j$ is matched to the design variance of $\Lambda$, $\sigma_{W_j}^2 = (q-1)^2\sigma_{\text{MAX}}^2 - \beta_j^2\sigma_j^2$. The decoder makes an estimate of the sum $T_1 + T_2 + \cdots + T_J$ and removes the common randomness variables, $W_j$, $j = 1, 2, \ldots, J$, to get an estimate of $U$ at the desired distortion. Note that as this works in expectation over the $W_j$, there exist fixed constants $w_1, w_2, \ldots, w_J$ that can serve the same role. □

We are now ready to prove our main theorem.

*Proof of Theorem 9:* Choose $q$ to be a prime number such that $q > L$ where $L$ is the number of receivers. We will use the channel network to convey Gaussian sources of length $k$. We will then connect the distortion performance back to sending bits using Lemma 5. Construct a new point-to-point channel network using $\mathcal{G}' = (\mathcal{V}', \mathcal{E}')$ as in the statement of the theorem. Let $C'$ be the multicast capacity of the $\mathcal{G}'$ which is given by the usual max-flow min-cut characterization. We would like to show that for any $\delta, \epsilon > 0$, we can achieve a multicast rate $R = C' - \delta$ on the original network with average probability of error not exceeding $\epsilon$.

First, using Lemma 4, we find an algebraic network code for $\mathcal{G}'$ that can be used to achieve a rate $R = C' - \frac{\delta}{2}$. This basically involves breaking every channel into several capacity $\lambda$ channels through time-division where $\lambda > 0$ is small enough that the rate loss due to rounding error is negligible. Thus, for a chunk of channel uses we get a $\lambda$ rate bit pipe which we designate to carry exactly one linear function. Let $P_\lambda, N_\lambda > 0$ be chosen such that $\lambda = \frac{1}{2}\log\left(1 + \frac{P_\lambda}{N_\lambda}\right)$. The number of inputs to a node is clearly upper bounded by the number of $\lambda$ bit pipes in the network which itself is upper bounded by:

$$|\mathcal{E}_{\text{UPPER}}| = \left(\max_{e \in \mathcal{E}'}\left\lfloor \frac{C_e}{\lambda} \right\rfloor\right)|\mathcal{E}'| \tag{70}$$

which is a constant that does not depend on $n$.

Each node is thus sent a finite number of functions, $U_1^k, U_2^k, \ldots, U_J^k$. Assume that each of these have variance at most $\sigma_U^2$. It makes MMSE estimates of these and prepares a new function, $V = \beta_1 U_1 + \beta_2 U_2 + \cdots + \beta_J U_J$ with $\beta_j \in \{0, 1, \ldots, q-1\}$ for each outgoing chunk of channel uses according to the algebraic network code. This function can be transmitted to the receiver with distortion at worst:

$$D_\ell = |\mathcal{E}_{\text{UPPER}}|(q-1)^2\sigma_U^2\left(\frac{N_\lambda}{N_\lambda + P_\lambda}\right)^\ell. \tag{71}$$

according to Lemma 6. Note that this holds whether the node in the new network is an actual node in the original network or a MAC. Thus, the processing at a node increases the distortion by at most a factor $|\mathcal{E}_{\text{UPPER}}|(q-1)^2$.

Let $\gamma = \frac{C' - \frac{\delta}{2}}{\lambda}$ which can be made an integer by choosing $\lambda$ appropriately. At the source we will create $\gamma$ i.i.d. Gaussian sources $S_1^k, S_2^k, \ldots, S_\gamma^k$ of length $k$ and variance 1. These will be relayed to the receivers by means of the algebraic network code and the coding method described above. The receiver will see functions of these original sources at some distortions. First note that the distortions will not exceed $(|\mathcal{V}_N| + |\mathcal{V}_{\text{MAC}}|)|\mathcal{E}_{\text{UPPER}}|(q-1)^2$. This is just the number of processing nodes multiplied by the maximum increase factor due to processing at one node.

The functions of the sources seen at the decoders can be written as a matrix transformation just as the original algebraic network code description. If the algebraic network code induces a transform $\mathbf{A}$ over $\mathbb{F}$ on the sources to a given receiver then the transform for Gaussian case is given by $\tilde{\mathbf{A}}$ which has the same entries as $\mathbf{A}$ but operations are over the reals. Since we assume $\mathbf{A}$ is full rank then $\tilde{\mathbf{A}}$ is full rank as well. Thus, we can solve for each original source at every receiver at distortion

$$D_\ell = \alpha\left(\frac{N_\lambda}{N_\lambda + P_\lambda}\right)^\ell \tag{72}$$

$$\alpha = \gamma^2(|\mathcal{V}_N| + |\mathcal{V}_{\text{MAC}}|)|\mathcal{E}_{\text{UPPER}}|(q-1)^2. \tag{73}$$

Finally, we invoke Lemma 5 to get a multicast rate from these distortions. We get that we can achieve any multicast rate satisfying:

$$R < \frac{\gamma}{2\ell} \log \left( \alpha \left( \frac{N_\lambda + P_\lambda}{N_\lambda} \right)^\ell \right) \tag{74}$$

$$= \frac{\gamma}{2} \log \left( 1 + \frac{P_\lambda}{N_\lambda} \right) - \frac{\gamma}{2\ell} \log \alpha \tag{75}$$

$$= \gamma \lambda - \frac{\gamma}{2\ell} \log \alpha \tag{76}$$

$$= C' - \frac{\delta}{2} - \frac{\gamma}{2\ell} \log \alpha \tag{77}$$

Choose $\ell$ large enough such that we can achieve $R = C' - \delta$. By making all the appropriate blocklengths large enough, we can make the probability of error arbitrarily small. This completes the proof. $\square$

## Acknowledgments

The authors would like to thank G. Bresler, U. Erez, K. Narayanan, A. B. Wagner and M. P. Wilson for valuable discussions. The work of M. Gastpar was supported by the National Science Foundation under CAREER Grant CCF-0347298 and the work of B. Nazer was supported by a National Science Foundation Graduate Research Fellowship.